\documentclass[runningheads]{llncs}
\usepackage{graphicx}
\usepackage{multirow}
\usepackage{biblatex}  
\begin{document}
\title{Uncertainty Quantification in Medical Image Segmentation with Multi-decoder U-Net }
%
%

\author{Yanwu Yang\inst{1,2} \and
Xutao Guo \inst{1} \and
Yiwei Pan \inst{1} \and
Pengcheng Shi  \inst{1} \and
Haiyan Lv\inst{5} \and
Ting Ma\inst{1,2,3,4}}

\institute{Department of Electronic and Information Engineering, Harbin Institute of Technology at Shenzhen, China 
\and
Peng Cheng Lab, shenzhen, China
\and
Advanced Innovation Center for Human Brain Protection, Capital Medical University, Beijing, China
\and
National Clinical Research Center for Geriatric Disorders, Xuanwu Hospital Capital Medical University, Beijing, China \and
MindsGo Co.Ltd, Shenzhen, China
}

\maketitle              
\begin{abstract}
Accurate medical image segmentation is crucial for diagnosis and analysis. However, the models without calibrated uncertainty estimates might lead to errors in downstream analysis and exhibit low levels of robustness. Estimating the uncertainty in the measurement is vital to making definite, informed conclusions. Especially, it is difficult to make accurate predictions on ambiguous areas and focus boundaries for both models and radiologists, even harder to reach a consensus with multiple annotations. In this work, the uncertainty under these areas is studied, which introduces significant information with anatomical structure and is as important as segmentation performance. We exploit the medical image segmentation uncertainty quantification by measuring segmentation performance with multiple annotations in a supervised learning manner and propose a U-Net based architecture with multiple decoders, where the image representation is encoded with the same encoder, and segmentation referring to each annotation is estimated with multiple decoders. Nevertheless, a cross loss function is proposed for bridging the gap between different branches. The proposed architecture is trained in an end-to-end manner and able to improve predictive uncertainty estimates. The model achieves comparable performance with fewer parameters to the integrated training model that ranked the runner-up in the MICCAI-QUBIQ 2020 challenge.

\end{abstract}
\section{Introduction}
Medical imaging segmentation plays a key role in diagnosis, monitoring, and treatment planning of the disease.
In recent years, segmentation in medical imaging achieves the state of the art with deep learning, at the same time saving physicians time and providing an accurate reproducible solution for diagnosis analysis and monitoring, even outperforming those conducted by doctors. 
However, despite high performance in segmentation accuracy, the modeling uncertainty is as important as accuracy, especially in medical scenarios\cite{hu2019supervised}.
In clinical experiments, there are biases in annotation among different doctors due to experience, understanding and so on.
Although the annotation would achieve a dice score over 0.9, the uncertainty still troubles. It might be also contradictory only to improve accuracy while ignoring this uncertainty.

Uncertainty measures are a promising direction since uncertainties can provide information as to how confident the system was on performing a given task on a given patient. This information in turn can be used to leverage the decision-making process of a user, as well as to enable time-effective corrections of computer results by for instance, focusing on areas of high uncertainty\cite{jungo2019assessing}.

Recent modules focusing on segmentation uncertainty were built based on probabilistic models, such as Bayesian neural network \cite{kendall2015bayesian}, Probabilistic U-Net \cite{kohl2018probabilistic} and so on \cite{gustafsson2020evaluating, kendall2015bayesian, kendall2017uncertainties, gal2016dropout}. These methods pinpoint the probabilistic of each pixel within the segmentation. Besides, ensemble models \cite{ronneberger2015u, lakshminarayanan2016simple, wenzel2020hyperparameter} would also obtain a considerable result to some extent, with small and simple models. However, these studies measure the uncertainty with the predictions’ discrepancies in various forms, which is not intuitive and there is no exact correspondence in the image \cite{wang2021medical}. Instead, the MICCAI-QUBIQ \footnote[1]{https://qubiq.grand-challenge.org/} challenge proposed to use a average dice loss to measure these uncertain areas with between 3 and 7 annotations from experts, where the uncertainty is quantified with labeling and would be analyzed in a reasonable way.

In this paper, we built a U-Net based architecture with multi-decoder for medical image segmentation uncertainty quantification, exploiting all the annotations as to increase the performance of segmentation and decrease the predictive uncertainty.
The contributions can be concludes as: 1) We built an end-to end architecture with multi-decoders for quantifying medical image segmentation uncertainty, where each decoder was implemented for measuring and fitting one annotation. 2) A cross loss function was carried out for optimizing, where the up-sampling information within different decoders were combined. Moreover, an auxiliary loss was carried out for improving performance. 3) Our architecture could remain high segmentation performance, at the same time measuring segmentation uncertainty. And finally we ranked second in the MICCAI-QUBIQ 2020 challenge.

\section{Method}

\begin{figure}
\includegraphics[width=\textwidth]{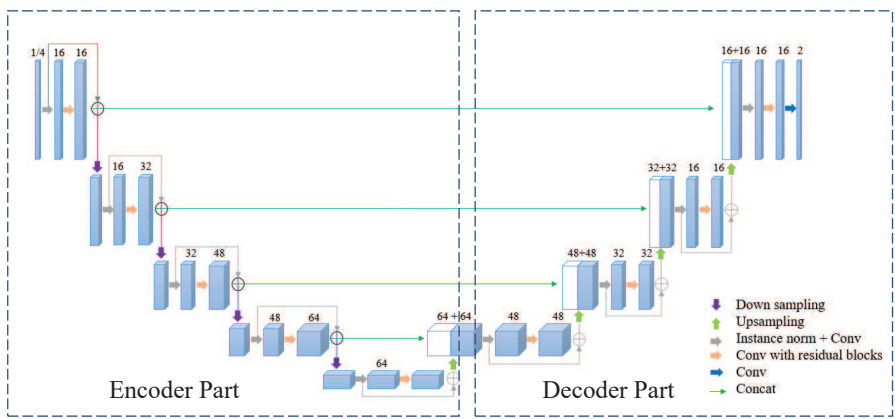}
\caption{The U-Net architecture implemented in our model, modified with residual blocks and instance normalization, where one decoder is shown in the figure. Our model is built upon this, with multiple decoders.} \label{fig1}
\end{figure}

\subsection{Neural network backbone}
Our architecture is built based on U-Net, as shown in Fig.~\ref{fig1}, where the channels of each stage are [16, 32, 48, 64, 64] respectively. Similarly, the context aggregation pathway is also implemented, which recombines multi-scale representations with shallower features to precisely localize the structures of interest. 
Each stage consists of two convolution groups with a residual structure. In detail, the convolution group is comprised of convolution operations with kernel size of 3, step size of 1, a relu activation function, and normalization. The upsampling is implmented by interpolation. 
The residual block was used for each stage, promoting the transmission of features in the network, and alleviating the gradient vanishing problem, making the network easier to train \cite{qiu2019embedded}. The transmitted features are combined via element-wise summation.
We further replace the batch normalization with instance normalization, which takes place in the feature space, therefore it should have more profound impacts than a simple contrast normalization in the pixel space \cite{huang2017arbitrary}.

\subsection{Mutli-decoder U-Net}
\begin{figure}
\includegraphics[width=\textwidth]{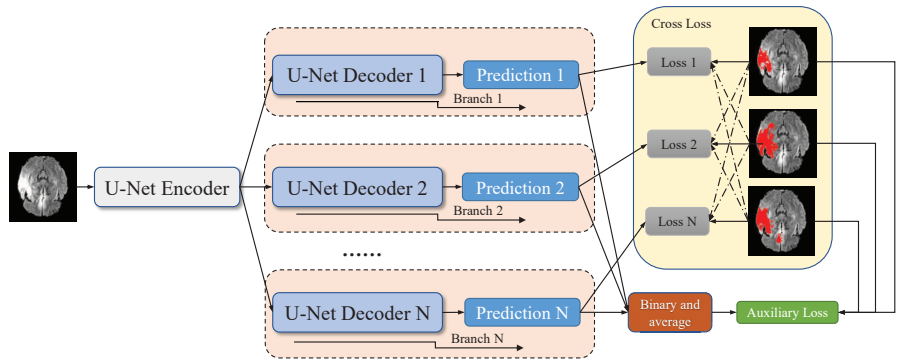}
\caption{The whole architecture of our multi-decoder U-Net, where each decoder is designed for predicting one annotation, and an auxiliary loss is implemented for improving training. The dotted line exhibit the training procedure, where the default dice loss is used as initialization, and the cross loss is enabled after a few epochs.} \label{fig2}
\end{figure}

Furthermore, multiple decoders are implemented in the architecture, where each decoder upsamples the same image representation, as shown in Fig.~\ref{fig2}. Within the framework, multiple annotations were measured integrally at the same time, and the framework received one input and multiple labels.The framework aims to increase the predict of each branch, at the same time improve the average dice of all the metric.

A cross loss function is proposed, aiming to orient the segmentation representation with context of multiple annotations. Specially, the loss of each decoder was a combination of dice loss and cross entropy loss.
\begin{equation}
L_{loss}^{i} = \alpha * L_{ce}^{i,i} + L_{dc}^{i,i} + \frac{1}{n - 1} \sum_{j,j \neq i}^{n} \beta_{j} * L_{dc}^{i,j}
\end{equation}
\begin{equation}
L_{dc}^{i,j} = 1 - \frac{2}{|K|} \sum_{k\in K}\frac{\sum^{I}_{i,j} u_{i,k}* v_{j,k}}{\sum_{i}^{I}u_{i,k} + \sum_{j}^{I}v_{j,k}}
\end{equation}
\begin{equation}
L_{ce}^{i,j} = -\sum_{k\in K}u_{i,k} * log(v_{j,k})
\end{equation}
where $u_{dc}^{i,j}$ denotes the average dice loss measured by the softmax output $u_{i}$ of the $i^{th}$ branch, the $j^{th}$ denotes one hot encoding of the ground truth segmentation map $v_{j}$, $j\in n$ represents the number of the annotations, and $u_{ce}^{i,j}$ denotes the cross entrophy loss in the same way. Both $u$ and $v$ have the same shape $I$ by $K$, $k\in K$ being the classes.

In addition, an auxiliary loss that is not differentiable, is adapted for assistant training, obtained by the dice coefficient between average predictions and averaged ground truths:

\begin{equation}\label{auxloss}
L_{ad} = -\frac{1}{10}\sum_{\tau = 0}^{1}{dice(Mask(i, \tau), Mask(j, \tau))}
\end{equation}
, where $\tau$ denotes the threshold ranging from 0 to 1 by step of 0.1, and $Mask$ denotes the operation to binarize the segmentation.

\section{Experiments and Results}
\paragraph{\bfseries{Dataset}}
The MICCAI-QUBIQ 2020 challenge provides 7 binary segmentation tasks in four different CT and MR data sets, where each task is segmented between three and seven times by different experts. In detail, the cohort includes 39 cases with 7 annotations for brain growth, 32 cases with 3 annotations for brain tumor, 24 cases with 3 annotations kidney, and 55 cases with 6 annotations for prostate. In total, there are seven binary segmentation tasks, two for prostate segmentation, one for brain growth segmentation, three for brain tumor segmentation, and one for kidney segmentation. Following the challenge, 48 cases of the prostate data, 34 cases of the brain growth data, 28 cases of the brain tumor data, and 20 cases of the kidney data are utilized as the training data. The remaining are the testing data. In addition, the data shapes within a single dataset are various, and are further padded to the same shape.

\begin{figure}
\includegraphics[width=\textwidth]{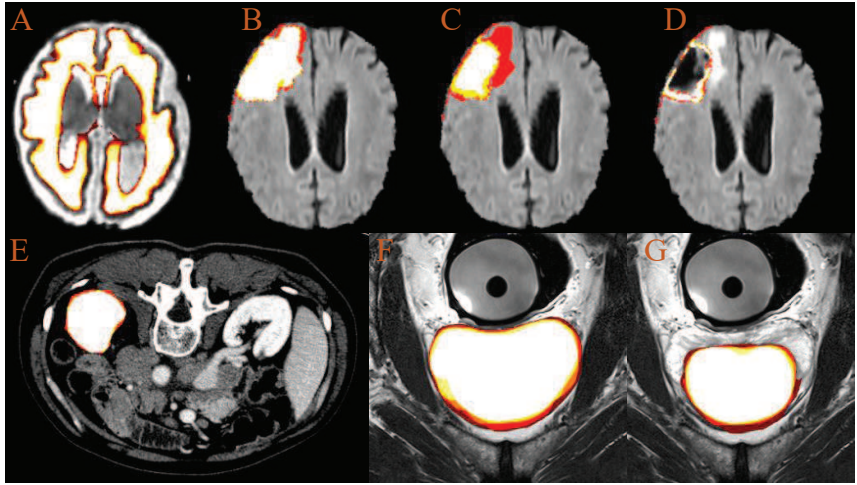}
\caption{Examples of the ambiguity in the dataset. A) The brain-growth task. B) The brain tumor task 1. C) The brain tumor task 2. D) The brain tumor task 3. E) The kidney task. F) The prostate task 1. G) The prostate task 2. The yellow and red areas introduce varying degrees of the uncertainty.} \label{figdata}
\end{figure}

\paragraph{\bfseries{Preprocessing}}
Normalization plays a key role in data representation, specially for data acquired with various protocols. For MRI images, we conducted the z-score normalization by subtracting the mean and dividing by the standard deviation. In the CT image data, we localized the regions of interests, filtering non-sensitive parts and subsequently rescale to [0, 1]. Considering that the annotations came from different experts and were labeled with inconsistent rule, we relabeled the annotations by reference to the overlap of raters. For example, the tumor task includes three annotations, and the three annotations would be averaged with values of [0.00, 0.33, 0.67, 1.00] and further binarized with threshold values of [0.33, 0.67, 1.00] into three new labels. 

\paragraph{\bfseries{Experiments}}
The result was conducted by the average prediction of each branch, which simulates experts to annotate the image. In this experiment, the performance is evaluated using the averaged dice score, measuring the uncertainty by binary at ten probability levels (that are 0.0, 0.1, 0.2, ..., 0.8, 0.9), as in (\ref{auxloss}).

However, the results would be seriously impacted when the convergence rate of all the branches vary, leading to non-uniform decreasing trends of losses. In terms of this, pretrained models were conducted, where all the models were trained from scratch to the same extent separately. Moreover, these models would be fine-tuned integrally. It can be estimated that although this balances the convergences of different branches, the integrated results are also impacted. In terms of this, the coefficients $\beta_{i}$ have been modified according to the values of losses from pretrained models in the integrated training progression.

In this experiment, no external data was used and the networks were trained from scratch. We conducted the experiments using Pytorch on one Nvidia V100 GPU. The learning rate is increased to $3e-4$ in 10 epochs using a linear warmup strategy. An adaptive moment estimate, Adam \cite{kingma2014adam} with weight decay of $1e-5$ is used as our optimizer.

\begin{table}
\caption{\centering Results in comparison with different methods.}\label{tab1}
\centering
\setlength{\tabcolsep}{5mm}
\begin{tabular}{c|cccc}
\hline
\multicolumn{2}{c}{Models} & U-Net & Ours & ensemble models\\
\hline
\multicolumn{2}{c}{average score} & 0.669 & 0.720 & {\bfseries 0.743}\\
\hline
\multicolumn{2}{c}{Brain-growth} & {\bfseries 0.483} & 0.477 &  0.481\\
\hline
\multirow{3}{*}{Brain-tumor} & task01 & 0.821 & 0.852 & {\bfseries 0.854} \\
& task02 & 0.814 & 0.871 & {\bfseries 0.898}\\
& task03 & 0.825 & 0.832 & {\bfseries 0.851}\\
\hline
\multicolumn{2}{c}{Kidney} & 0.845 & 0.870 & {\bfseries 0.872}\\
\hline
\multirow{2}{*}{Prostate} & task01&0.472 & 0.553 & {\bfseries 0.554}\\
& task02 & 0.423 & 0.584 & {\bfseries 0.689}\\
\hline
\end{tabular}
\end{table}

In terms of this novel task using multiple annotations and the average dice loss for measurement, related models are few and could hardly be implemented to achieve considerable performance. Here we designed a baseline, where several models are trained for multiple annotations. The final results were averaged with the models in the same way. Furthermore, we trained the network three times by modifying hyper-parameter of $\alpha$ and $\beta_{i}$ as an ensemble to predict the respective validation and improve the 
generalization ability of the model. Validation set results were evaluated using the online evaluation platforms to ensure comparability with other participants. 

\paragraph{\bfseries{Results}}
\begin{figure}
\includegraphics[width=\textwidth]{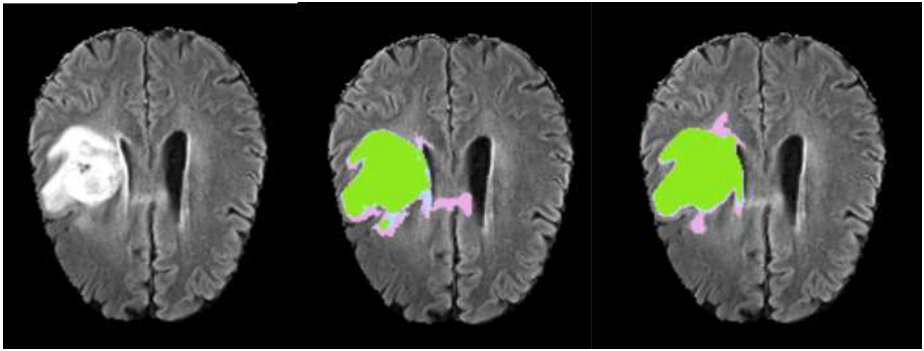}
\caption{Brain tumor segmentation results in comparison, which is shown as ground truth, averaged annotations and our results from left to right. The colors represent the uncertainty of the annotation and the segmentation.} \label{fig3}
\end{figure}

The results are shown in Table.~\ref{tab1}. Our proposed multi-decoder U-Net outperforms the U-Net baseline in six tasks, with 0.852 in brain tumor task1, 0.871 in brain tumor task2, 0.832 in brain tumor task3, 0.870 in kidney, 0.553 in prostate task1, and 0584. in prostate task2. Besides, the ensemble models performs the best. While, all the models perform badly in the brain-growth task, mainly because that the result is corresponding to the number of annotations and it is more likely that the gap between the annotations is enhanced inevitably.

\section{Conclusion}
In this work, we proposed a multi-decoder U-Net to calibrate the uncertainty with multiple annotations in an end-to-end manner. The novel task measuring the segmentation with multiple annotations and the averaged dice loss brings a way to further analyze and quantify the uncertainty, which has a better ability to reveal tiny and ambiguous structures and has a potential for diagnosis. Our method is simple and easy to be facilitated, where the decoder imitates the experts for labeling. The intuitive training strategy outperforms other complex models and achieves competitive results on six tasks. In order to perform a stable training, we fine-tuned the models using weights of baseline models and relabeled the labels by averaging and binarizing the annotations. The combination of the proposed model and training strategy makes contribution to the results. As future work, we would like to optimize the boundaries using multiple annotations and propose an adaptive segmentation model with uncertainty quantification.
\section{Acknowledge}

This study is supported by grants from the National Key Research and Development Program of China (2018YFC1312000) and Basic Research Foundation of Shenzhen Science and Technology Stable Support Program (GXWD20201230155-427003 - 20200822115709001)
\printbibliography
\end{document}